\documentclass[11pt,twoside]{article}

\usepackage{asp2006}
\usepackage{epsf}
\usepackage{psfig}
\usepackage{lscape}
\usepackage{graphicx}
\def\ltsim{\; \raise0.3ex\hbox{$<$\kern-0.75em \raise-1.1ex\hbox{$\sim$}}\; }
\def\gtsim{\; \raise0.3ex\hbox{$>$\kern-0.75em \raise-1.1ex\hbox{$\sim$}}\; }
\def\ie{{\it i.e.,~}}

\def\eg{{\it e.g.,~}}
\begin{document}
\title{Large-scale diffuse radio emission from clusters of galaxies and the importance of low frequency radio observations}   
\author{Rossella Cassano}   
\affil{INAF -- Istituto di Radioastronomia, Bologna, Italy}    

\begin{abstract} 
The presence of non-thermal components in galaxy clusters is now clearly established. Diffuse radio emission from the Intra Cluster Medium (ICM) of several galaxy clusters is revealed in the form of
radio halos and relics. These emissions are synchrotron radiation from a population of relativistic electrons mixed with the thermal gas and diffusing through $\approx\,\mu G$ turbulent magnetic fields.
Radio Halos are surely the most interesting evidences of cluster non-thermal activity and understanding their origin is one of the most intriguing problems of the physics of the ICM. I review observational and theoretical results obtained in the last few years and discuss the impact of present (e.g. GMRT) and future low frequency radiotelescopes (LOFAR, LWA) in our understanding of non-thermal phenomena in galaxy clusters.
\end{abstract}


\section{Introduction}   
\label{Intro}

Galaxy clusters have a key role in the cosmic hierarchy 
as they are the largest bound structures in the Universe.
They extend over $\approx 2-4$ Mpc and contain $\approx 10^{14}-10^{15}\,M_{\odot}$ 
of hot ($10^{8}$K) gas ($\sim15-20\%$), galaxies ($\sim10\%$) 
and dark matter ($\sim70\%$).
There is now general agreement on a hierarchical picture for the formation
of cosmic structures, in which galaxy clusters are supposed to form by 
accretion of matter and merging between smaller units at the intersection
of filaments which form the ``cosmic web'' (\eg Borgani et al. 2004).
During mergers, shocks are driven by the gravity of the dark matter
in the diffuse barionic component, which is heated up to the observed 
temperature.

Galaxy clusters are bright X-ray sources with X-ray luminosities of $\sim 10^{43}-10^{45}$ erg/s, 
due to thermal Bremsstrahlung radiation from the hot gas.
Radio observations have discovered an increasing number of Mpc-sized emissions from the
Intra Cluster Medium (ICM), Radio Halos, at the cluster center, and Radio Relics, at the cluster 
periphery (\eg Feretti 2003; Ferrari et al. 2008). These sources are due to synchrotron emission 
from ultra relativistic electrons diffusing through $\mu$G turbulent magnetic fields. 
Such non-thermal components mixed with the thermal ICM 
may drive still unexplored physical processes and this can modify our simplified
view  of the ICM (Schekochihin et al. 2005; Subramanian et al. 2006; Brunetti \& Lazarian 2007;
Guo et al. 2008).

Cluster mergers are believed to be the most important sources of non-thermal components 
in galaxy clusters: a fraction of the energy dissipated during these mergers 
could be channelled into the amplification of the magnetic 
fields (\eg Dolag et al. 2002; Br\"uggen et al. 2005; Subramanian et al. 2006; Ryu et al. 2008)
and into the acceleration of high energy particles via shocks and turbulence 
(\eg En\ss lin et al. 1998; Sarazin 1999; Blasi 2001; Brunetti et al. 2001, 2004; Petrosian 2001;
Miniati et al. 2001; Fujita et al. 2003; Ryu et al. 2003; Hoeft \& Br\"uggen 2007; Brunetti \& Lazarian 2007; Pfrommer et al. 2008, Brunetti et al. 2009).

Large scale diffuse radio emission from galaxy clusters, in the form of radio halos and relics, 
has steep spectrum (typical spectral indices are $\alpha\approx 1.2-1.3$, with $F(\nu)\propto\nu^{-\alpha}$) and this makes low frequency radio observations ideal tools 
to study these sources, their origin and evolution.
The recent discovery of very steep spectrum diffuse emission in the galaxy cluster Abell 521
provides a glimpse of what the next generation of radiotelescopes such as LOFAR \& LWA might find 
in galaxy clusters (Brunetti et al. 2008).

Radio halos are the most spectacular diffuse synchrotron sources and are the main focus of 
this review, that is organized as follows: Sect.2 summarizes relevant observational results 
obtained in the last few year on the subject; I focus on the physics of relativistic particles in clusters in Sect.3 and on the origin of radio halos in Sect.4. I report on the expected statistical properties of radio halos at high and low frequency in Sect.~5, conclusions are given in Sect.6.

\section{Observations}
\label{Obs}

\begin{figure}
\includegraphics[width=6.3cm,height=6.3cm]{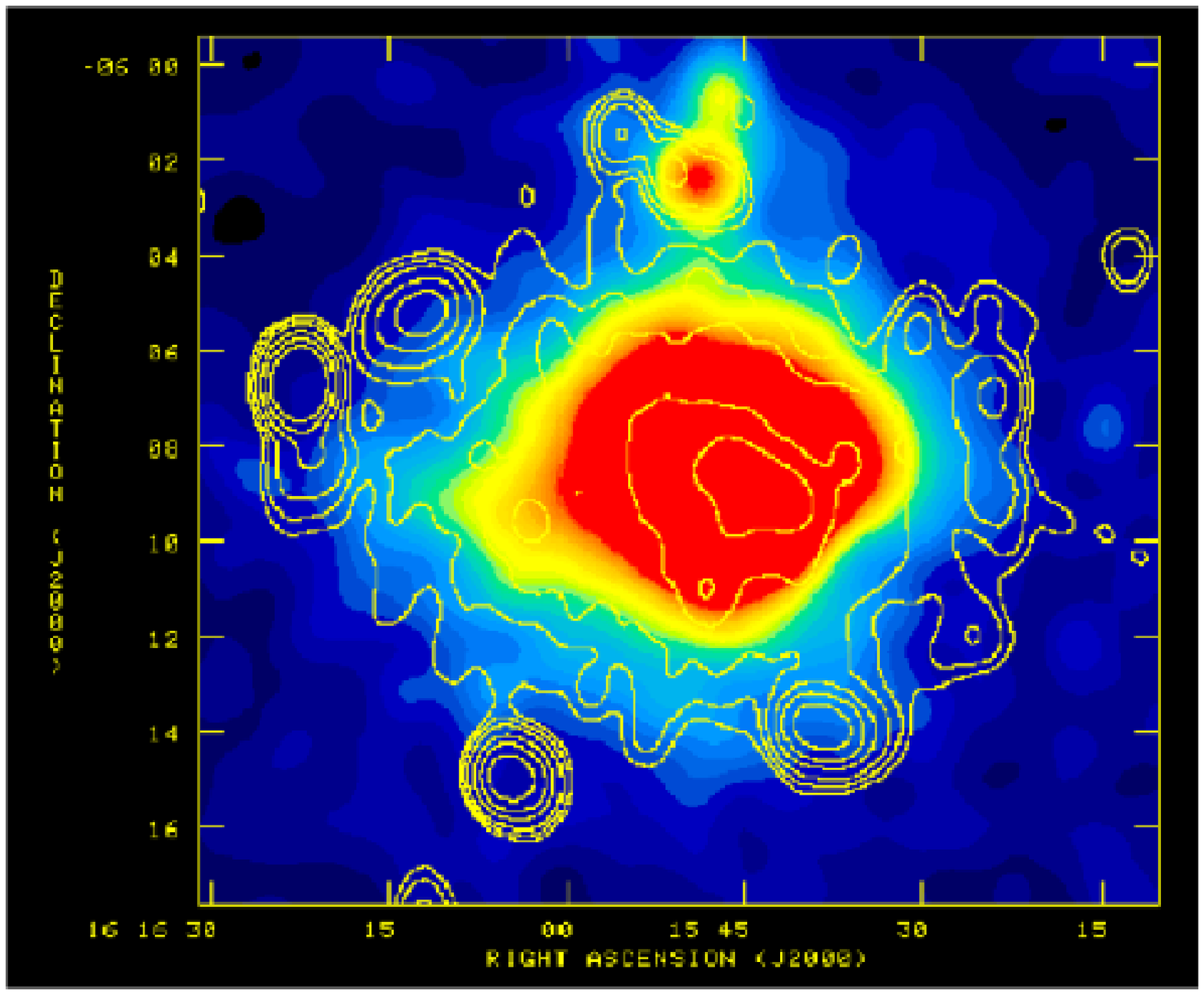}
\includegraphics[width=6.6cm,height=6cm]{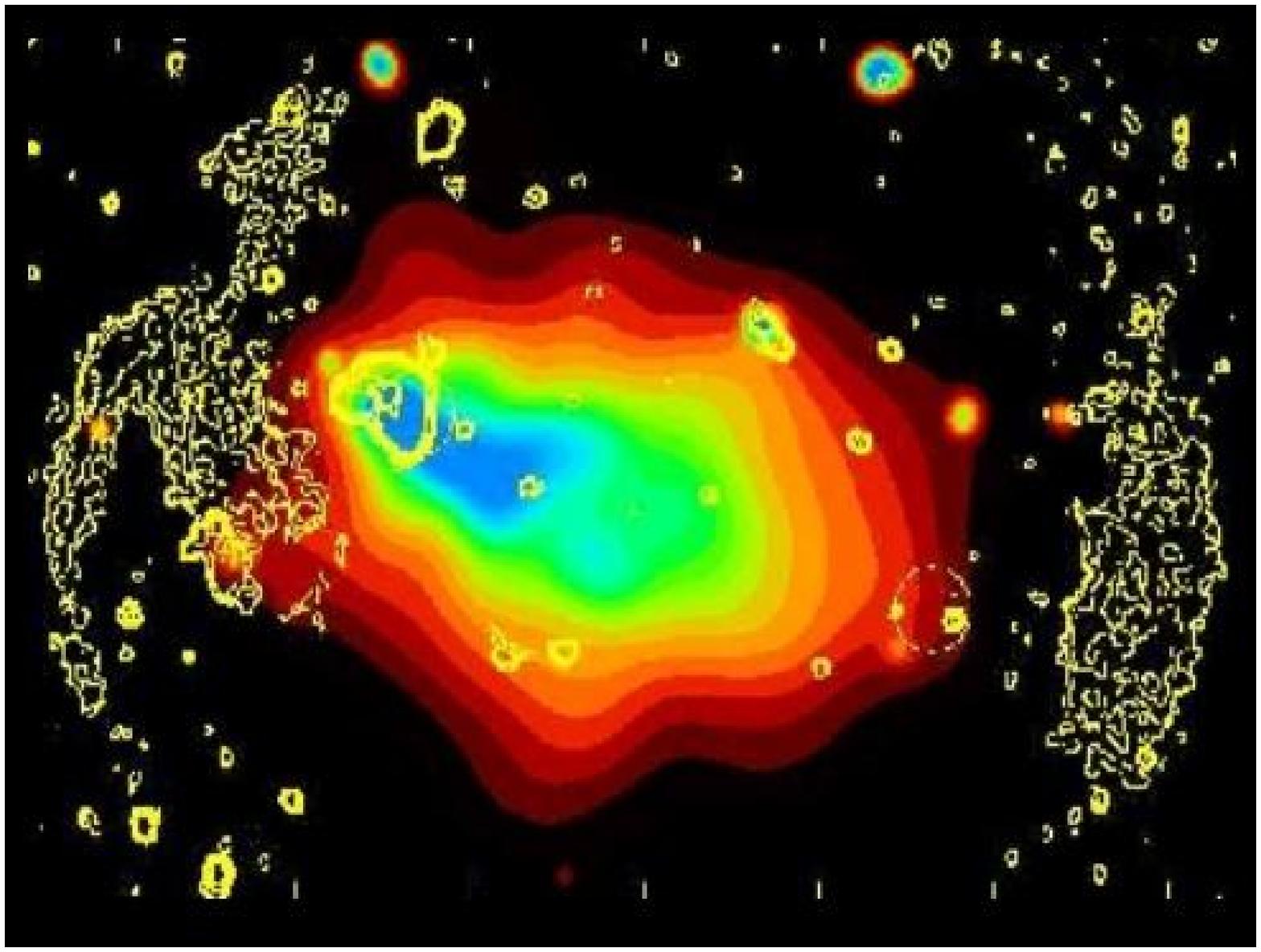}
\caption{{\bf a)} {\it Radio Halo} in Abell 2163: radio contours at 1.4 GHz overlaid
on the ROSAT X-ray emission (Feretti et al. 2001). {\bf b)} {\it Radio Relics} in the cluster Abell 3376:
radio contours at 1.4 GHz overlaid on the ROSAT X-ray emission (Bagchi et al. 2006).}
\label{fig:A2163}
\end{figure}

{\it Radio Halos} (RH hereafter) are defined in the literature as large diffuse non-thermal radio sources 
permeating the cluster centers which are not associated with any single active galaxy but 
rather with the diffuse ICM (Fig.\ref{fig:A2163}.a; Feretti 2003, Ferrari et al. 2008).
In general they have a regular shape, low surface brightness ($\sim \mu \mathrm{Jy/arcsec}^2$ at $1.4$ GHz), with typical luminosity of $\sim 5\cdot10^{23}-5\cdot10^{25}\,\rm{h_{70}^{-2}}$ Watt/Hz at $1.4$ GHz, and size of $\approx 1$ Mpc. They have a steep radio spectrum and low or negligible
polarization ($< 10\%$)\footnote{The only exception being the RH in Abell 2255 which shows a filamentary structure strongly polarized (Govoni et al. 2005).}.

RHs are difficult to detect because of their low surface brightness and due to the difficulty in separating the diffuse emission from the embedded discrete radio sources. Despite the observational difficulties, several studies were undertaken to identify candidate RHs in clusters at $z\leq 0.2$ from radio surveys (the NVSS at 1.4 GHz by Giovannini et al. 1999; the WENSS at 327 MHz by Kempner \& Sarazin 2001). Those studies suggested the rarity of such sources and their tendency to be located in 
X-ray luminous clusters (Giovannini et al. 1999), yet it remained unclear the role of selection biases due to the brightness limits of the used radio surveys (Kempner \& Sarazin 2001; Rudnick et al. 2005). 

\begin{figure}
\includegraphics[width=6.6cm,height=6.5cm]{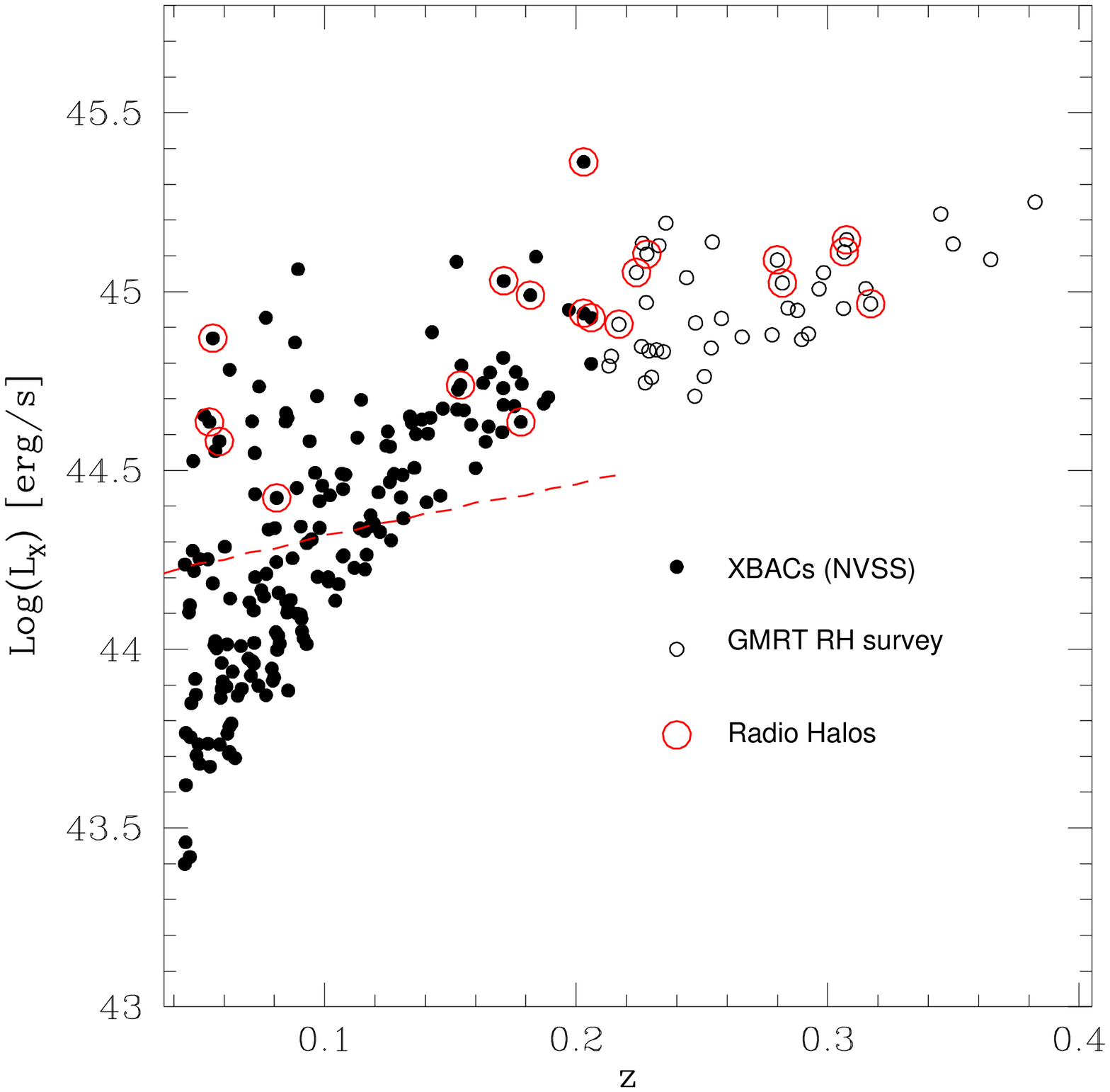}
\includegraphics[width=6.6cm,height=6.5cm]{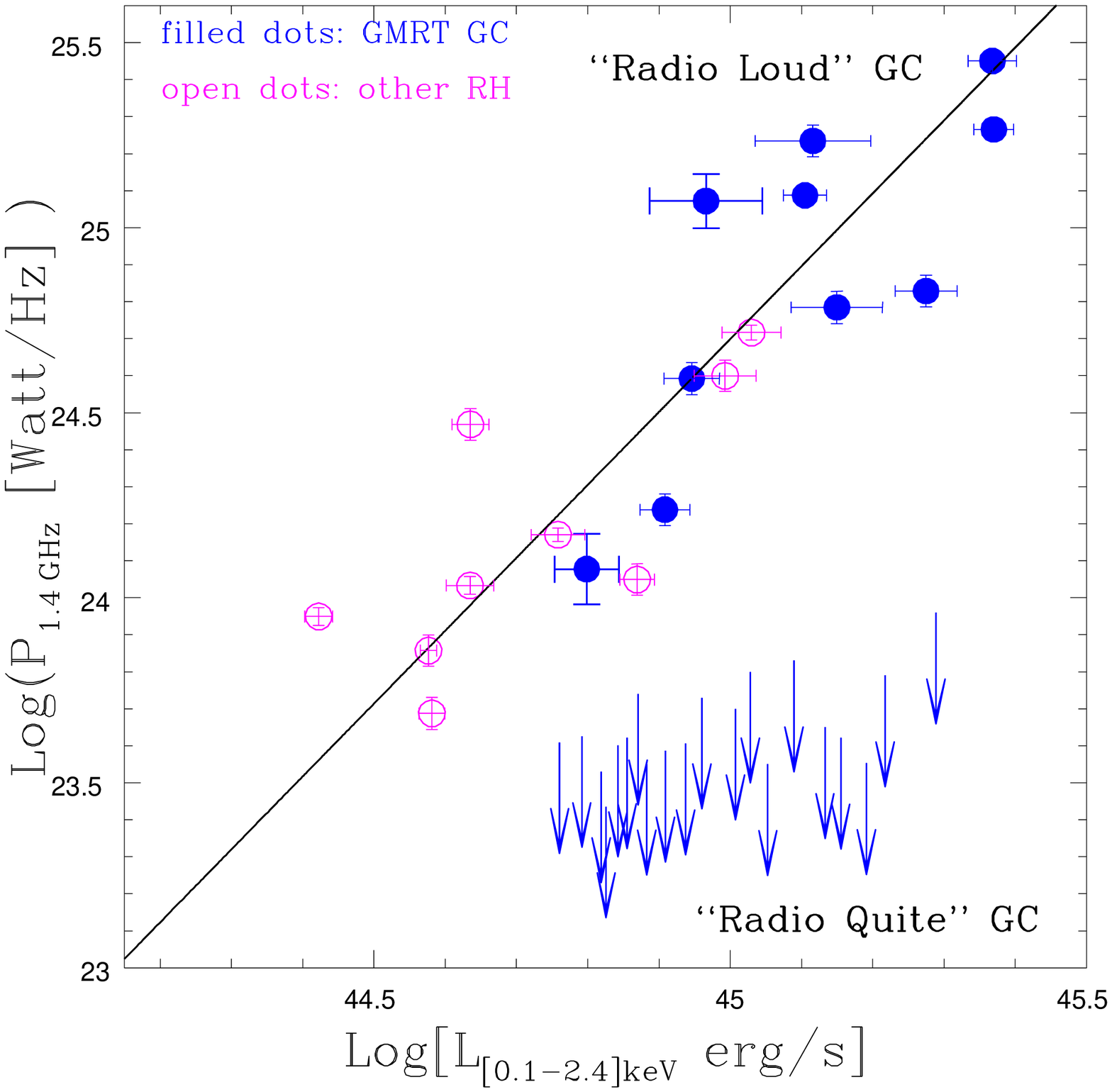}
\caption{{\bf a)} X-ray luminosity (0.1-2.4 keV) versus redshift for the sample of low redshift clusters (XBACs/NVSS, filled dots) and for the GMRT sample (open dots). RH clusters are marked with red big circles. The dashed line give an estimate of the minimum X-ray luminosity of clusters with RHs visible in the NVSS (\eg Cassano et al. 2008). 
{\bf b)} Distribution of galaxy clusters in the $P_{1.4}-L_X$ plane (adapted from Brunetti et al. 2007). Filled dots are clusters at $z\geq 0.2$ (from GMRT sample and from literature), empty dots are clusters at lower redshift reported to highlight the $ P_{1.4}-L_X$ correlation (solid line).
Upper limits (arrows) are GMRT clusters with no hint of cluster-scale emission and their distribution should be compared with that of clusters with similar redshift (filled dots).}
\label{Lx_Pr}
\end{figure}

In this respect the contribution of the GMRT has been important due to the ``GMRT RH survey'', dedicated to the search of RHs at 610 MHz in X-ray luminous ($L_X\geq 5\cdot10^{44}\,h_{70}^{-2}$) galaxy clusters at $0.2\leq z\leq 0.4$ (Venturi et al. 2007; Venturi et al. 2008, see also Venturi this volume).
These observations were specifically designated to avoid problems in the detection of cluster-scale  emission due to the missing of short-baselines in the interferometer observations and to image, at the same time, both compact and extended sources in the selected clusters.
Fig.\ref{Lx_Pr}.a shows the distribution of clusters observed with the GMRT (open dots) in the plane $L_X-z$ (together with clusters belonging to a sample at $z<0.2$, filled dots): RHs are only found in 30\% of clusters in the GMRT sample confirming that these sources are relatively {\it rare}.
Furthermore, the analysis of the distribution of RHs with the cluster X-ray luminosity (combining the GMRT sample with the low redshift sample) allows to derive that the fraction of clusters with RHs depends on the cluster X-ray luminosity (and mass), specifically: only $\sim10\%$ of clusters with $3\cdot 10^{44}-8\cdot 10^{44}$ erg/s host a RH, while $\sim40\%$ of clusters with $L_X\geq8\cdot 10^{44}$ erg/s have a RH (Cassano et al. 2008).
The important point of the ``GMRT RH survey'' is that the derived fractions of clusters with RHs are
reliable since they are not affected by the sensitivity of the observations. This can be immediately understood from Fig.\ref{Lx_Pr}.b that reports the distribution of clusters in the $P_{1.4}-L_X$ plane showing that the upper limits to the radio power in the case of clusters without RHs are one order of magnitude below the region of RHs (Brunetti et al. 2007; Venturi et al. 2008).

Another important point concerns the synchrotron spectrum of RHs, whose shape is still poorly known since RHs are generally observed only at a few frequencies.
The best studied among halos is Coma C in the Coma cluster (\eg Willson 1970; Schlickeiser et al. 1987; Giovannini et al. 1993; Thierbach et al. 2003).
The integrated radio spectrum of the Coma halo is a steep power-law with $\alpha\simeq1.2$
at frequency below 1.4 GHz, while observations at higher frequencies reveal the presence of
a cut-off (Schlickeiser et al. 1987; Thierbach et al. 2003) which is interpreted due to a break in
the spectrum of the emitting electrons.
Low frequency observations are very important for spectral studies since they allow to increase the frequency range. For instance, the spectrum of the small RH in Abell 3562 shows a break
when observed between 240 MHz and 1.4 GHz (Giacintucci et al. 2005). Remarkably, low frequency
radio observations allowed the discovery of the RH in A521 whose very steep spectrum ($\alpha\simeq 2$)
implies the presence of a high frequency cut-off (Brunetti et al 2008). In Sect.\ref{sec:picture} I discuss the importance of spectral and statistical studies to address the origin of RHs.

Even though this review is focused on RHs, for seek of completeness it is worth mentioning that 
several clusters are known to host {\it Radio Relics} (Fig.\ref{fig:A2163}.b) which are similar 
to RHs in terms of low surface brightness, large size, luminosity and steep spectrum, but in general are linearly polarized, elongated in shape and located in cluster peripheral regions (\eg Feretti 2003; Ferrari et al. 2008; see also Kale \& Dwarakanath this volume).
For these properties they are usually interpreted in terms of interection of shocks with thermal or ghost
plasma in the ICM. In particular two mechanisms to accelerate the radio-emitting
electrons have been proposed: diffuse shock acceleration (Fermi I) of thermal or fossil electrons
(En\ss lin et al. 1998; R\"ottiger et al. 1999) and adiabatic compression of fossil radio plasma by merger shock waves (En\ss lin \& Gopal-Krishna 2001; En\ss lin \& Br\"uggen 2002).

\section{Dynamics of relativistic particles in galaxy clusters}
\label{particles}

As discussed in Sects.~\ref{Intro} and \ref{Obs} RHs prove the presence
of non-thermal components such as magnetic field and relativistic particles
mixed with the thermal ICM. Understanding the energetics and physics of these
new components is important not only to draw the picture of the non-thermal phenomena
in galaxy clusters but also to understand how these components may eventually affect the physics
of the thermal ICM. In this Sect. I briefly discuss the origin of relativistic particles in clusters, 
while I refer the reader to the recent reviews by Govoni \& Feretti (2004) and by Dolag, Bykov \& Diaferio (2008) for the origin and evolution of magnetic fields.

Clusters host a large number of sources of cosmic rays: galaxies, AGN, powerful galactic winds
(\eg En\ss lin et al. 1997; V\"olk \& Atoyan 1999). In addition, cluster formation is also believed to
provide a contribution to the injection of cosmic rays in the ICM due to the formation of shocks which
may accelerate relativistic particles (Blasi 2001; Miniati et al. 2001; Fujita \& Sarazin 2001;
Gabici \& Blasi 2003; Ryu et al. 2003; Pfrommer et al. 2006; Hoeft \& Br\"uggen 2007; Vazza et al. 2008).

\begin{figure}
\centerline{\includegraphics[width=6.7cm,height=6.7cm]{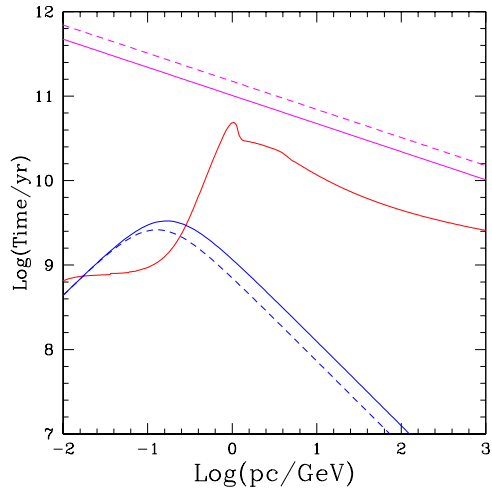}}
\caption{Time scale for energy losses of protons (middle) and electrons (bottom). The lines in the upper part give the diffusion time scale for a Kolmogorov spectrum of magnetic field fluctuations, for
a magnetic field of $1\mu$G (solid line) and $3\mu$G (dashed line). Credit: Blasi et al. (2007).}
\label{fig:losses}
\end{figure}

Once injected in the ICM the relativistic particles are subject to energy losses.
Relativistic {\it electrons} with momentum $p_e=m_ec\gamma$ lose energy through ionization losses and Coulomb collisions which dominate for $\gamma< 100$, and via synchrotron and inverse Compton scattering off the Cosmic Microwave Background (CMB) photons, which dominate at higher energies (\eg Sarazin 1999). On the other hand relativistic {\it protons} lose energy mainly through {\it pp} inelastic scattering, while Coulomb losses become important at lower energies. The time scales for losses due to the combination of these process (\ie the time after which particles lose a substantial fraction of their initial energy) are illustrated in Fig.~\ref{fig:losses}; the figure shows also the time scale necessary to the particles to diffuse out of galaxy clusters (magenta lines). The important point is that for the bulk of relativistic protons in galaxy clusters both the times cale for energy losses and the diffusion time scale are very large ($\approx$ Hubble time; Berezinsky et al. 1997; V\"olk et al. 1996). This implies that relativistic protons are accumulated in the ICM during the cluster lifetime. The confinement of relativistic protons enhances the probability to have {\it pp} collisions, that in turn give gamma ray emission via decay of neutral pions produced during these collisions and synchrotron emission (and IC emission) due to the production of secondary electrons.
Since protons accumulate in galaxy clusters, the emissions from their secondary products
can be roughly thought as a ``stationary'' signal.

On the other hand, relativistic electrons are short living particles that radiate their energy in the region
where they are produced (\eg Jaffe 1977). Specifically, electrons emitting synchrotron radiation around $\sim 1$ GHz have an energy of the order of $\approx 10\,B_{\mu G}^{1/2}$ GeV and a life-time of $\approx 10^{8}$ yr. During this timescale electrons can only diffuse for a few tens of kpc, which is very small compared with the observed size of RHs (Mpc scale). 
These arguments lead to the requirement that the electrons responsible for the observed radio emission in the form of RHs should be generated or accelerated everywhere in the cluster: either secondary electrons from {\it pp} collisions (Dennison 1980; Blasi \& Colafrancesco 1999), or electrons re-accelerated {\it in situ} through second order Fermi mechanisms by turbulence injected in the ICM during cluster mergers (Brunetti et al. 2001; Petrosian 2001).

\section{Origin of Radio Halos in galaxy clusters}
\label{sec:picture}

Extended and fairly regular diffuse synchrotron emission is expected from secondary electrons produced during {\it pp} collisions in the ICM; this has been proposed as a model for the origin of RHs
(\eg Dennison 1980; Blasi \& Colafrancesco 1999). In this context some level of $\gamma$-ray emission 
from secondary $\pi^0$ is unavoidably expected. As yet, upper limits only are obtained from present $\gamma$-ray observations of galaxy clusters (\eg Reimer et al. 2003; Aharonian et al et al. 2008). The spectrum of the RH recently discovered in A521 is inconsistent with a secondary origin of the emitting electrons due to simple energy arguments (Brunetti et al. 2008). Furthermore, the statistical properties
of RHs show a more complex behaviour than that expected from the secondary model, at least in its simplest form. Indeed, since all clusters have suffered mergers (hierarchical scenario) and protons are mostly 
confined within clusters, extended emission should be common in clusters, and the rarity of RHs, that is now well established from observations (Sect.~\ref{Obs}), is somewhat difficult to reconcile with this scenario.

Given these difficulties, a more complex model to explain the origin of RHs has been worked out in the last decade, the ``re-acceleration scenario'', in which fossil (and/or secondary) electrons are supposed to be stochastically re-accelerated (via second order Fermi mechanisms) due to the interaction with MHD turbulence injected in the ICM during cluster mergers (\eg Brunetti et al. 2001; Petrosian 2001). 
The historical motivations for this scenario was the connection observed between RHs and cluster mergers (\eg Buote 2001), and the cut-off at high frequency observed in the spectrum of the
Coma RH (\eg Schlickeiser et al. 1987). Indeed since stochastic particle acceleration is a poorly efficient 
process, due to the competition with energy losses electrons could be re-accelerated only up to a relatively low maximum energy ($\gamma\approx10^{4}$), implying an unavoidable high frequency cut-off in the radio spectrum of RHs.

At the same time, theoretically protons are believed to be the most important non-thermal components implying that the process of secondary electron production (at some level) is unavoidable and that 
the emerging general picture is very complex: a mixed population of relativistic particles (protons, secondary and primary electrons/positrons, re-accelerated particles) should coexist in the ICM together with turbulent magnetic fields and thermal particles. This implies a complex broad band non-thermal spectrum (from radio to gamma rays) from galaxy clusters that can be thought as composed by two main
components: a long-living one that is emitted by secondary particles (and by $\pi_0$ decay) continuously
 generated during {\it pp} collisions in the ICM, and a transient component that is due to the re-acceleration of relativistic particles by MHD turbulence generated (and then dissipated) in cluster mergers (Brunetti et al 2009). The first component is presently constrained by $\gamma$-ray observations of galaxy clusters and by radio observations of galaxy clusters without RHs (Brunetti et al. 2007), while the second component, that should be connected with cluster mergers, may explain RHs.

\section{Statistical properties of radio halos and low frequency observations}
\label{Sec:stat}

\begin{figure}
\includegraphics[width=7.3cm,height=6.2cm]{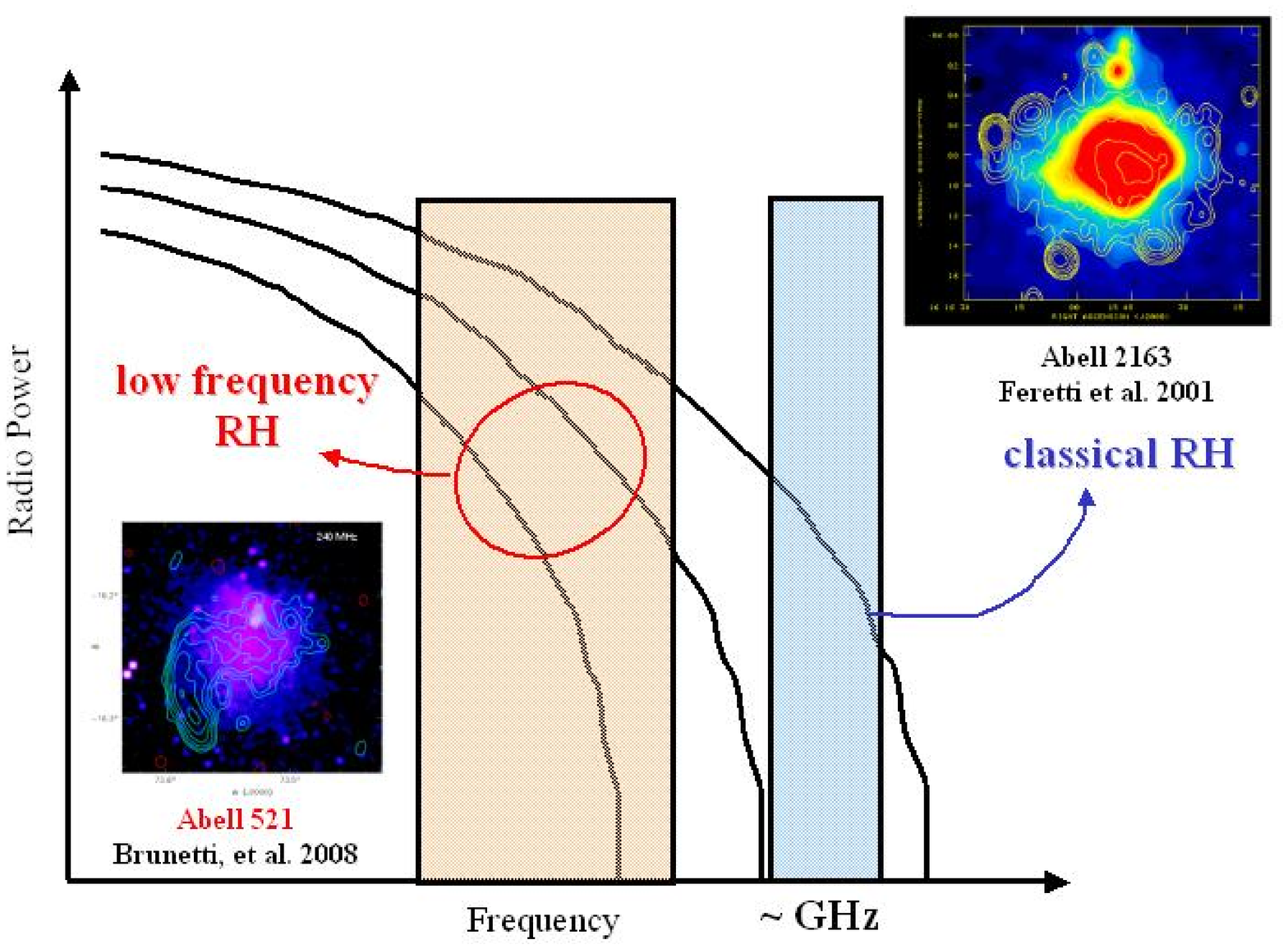}
\includegraphics[width=6.3cm,height=6.4cm]{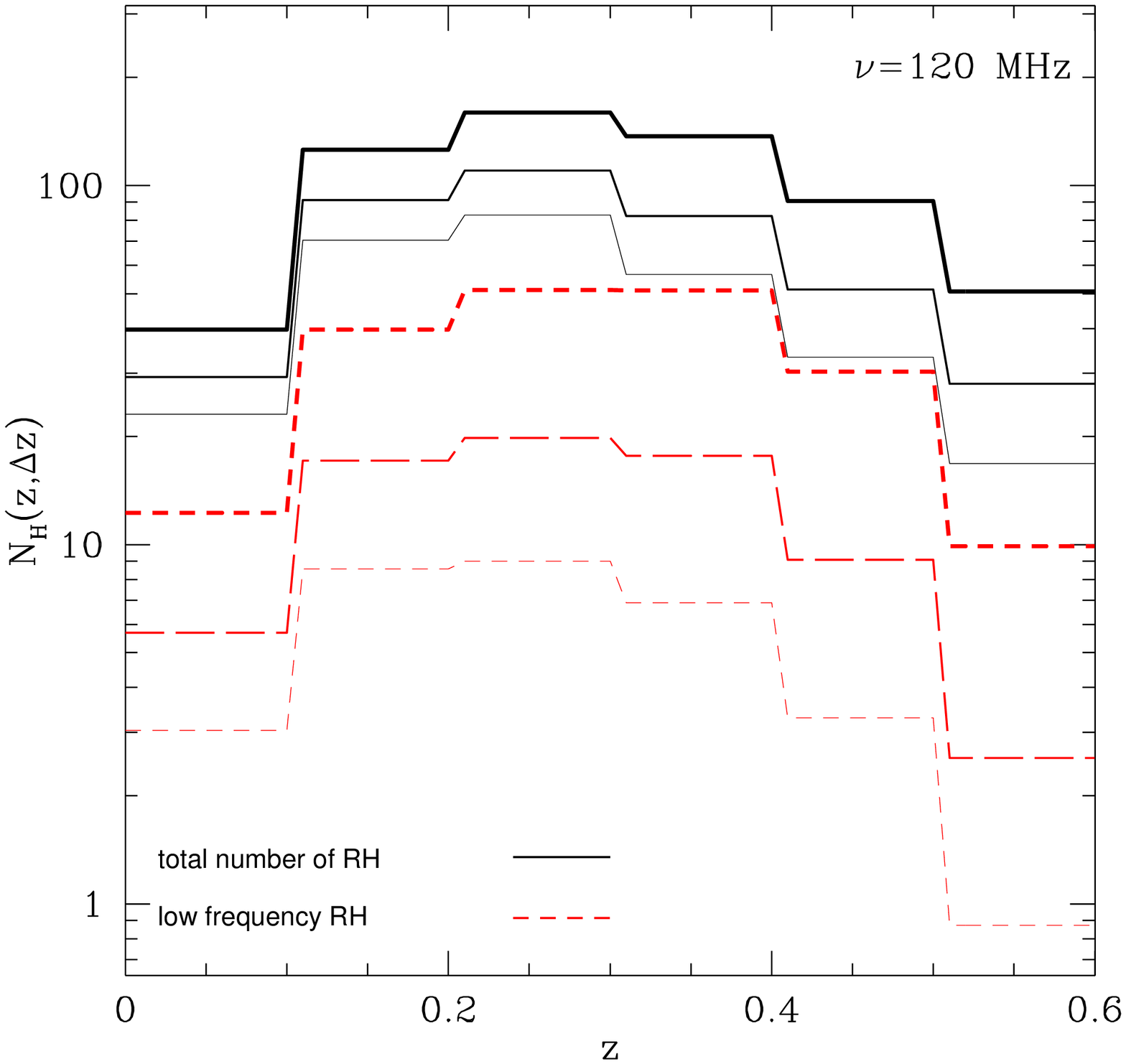}
\caption{{\bf a)} A schematic representation of the synchrotron radio spectra of ``classical'' RHs,
visible up to GHz frequencies, and of low frequency RHs, visible only at low radio frequencies. {\bf b)} Number (all sky) of RHs at 120 MHz as a function of redshift that can be expected 
at the sensitivity of LOFAR surveys. 
Calculations are performed assuming the following parameters: rms$=0.5, 1, 1.5$ mJy/beam
(from top thick lines, to bottom thin lines), and beam$=35'' \times 35''$. 
Solid lines give the total number of RHs at 120 MHz, while dashed lines give the number of low frequency RHs ($\nu_c<600$ MHz).} 
\label{fig:ussrh}
\end{figure}

The picture of the formation and evolution of RHs in galaxy clusters that comes out from the re-acceleration scenario is very complex, and tightly connected with the process of formation 
and evolution of the hosting clusters. In this context, the statistical properties of RHs depend on the interplay between the rate of cluster-cluster mergers in the Universe and the fraction of the energy dissipated during these mergers that is channeled into MHD turbulence and re-acceleration of high energy particles. In the last few years this connection has been investigated through Montecarlo procedures 
(Cassano \& Brunetti 2005; Cassano et al. 2006).
Despite our ignorance on many details of the physics of turbulence and of particle acceleration in the ICM, 
this model predicts some basic expectations for the statistical behaviour of the properties of RHs.
Only massive and merging clusters, where enough energy can potentially be channelled into particle re-acceleration, are expected to host RHs. In addition, since the turbulent
energy injected during mergers is expected to scale with the cluster thermal energy (Cassano \& Brunetti 2005; Vazza et al. 2006), the fraction of clusters with RHs is expected to increase with the cluster mass (or X-ray luminosity); remarkably this is in line with results from recent radio surveys (Cassano et al. 2008; Venturi et al. 2008).

Future low frequency radio observations are expected to shed new light on our understanding of RHs.
As a general point, regardless of their origin, a large number of RHs is expected at fainter radio fluxes
by simply considering the extrapolation of their 1.4 GHz number counts through the radio power--X-ray luminosity correlation (En\ss lin \& R\"ottgering 2002). 
These faint RHs are hardly detectable with present facilities at 1.4 GHz, but
since they have steep radio spectra they should appear more luminous at lower frequencies,
and thus LOFAR and LWA should discover a large number of these objects.

In addition to this general argument the re-acceleration scenario predicts that the fraction of clusters 
with RHs depends on the frequency of observations. 
This can be easily understood from Fig.\ref{fig:ussrh}.a where spectra of RHs are illustrated. The presence of a cut-off in the expected spectrum of RHs affects our capability to detect
these sources making it difficult to observe them at frequencies larger than the cut-off frequency, $\nu_c$. The value of $\nu_c$ is essentially determined by the acceleration efficiency,
which in turns depends on the flux of MHD turbulence dissipated in the re-acceleration of relativistic electrons. As a consequence, in this model it is expected that ``classical'' RHs, observed up to $\sim 1$ GHz, should originate in connection with the most powerful cluster-cluster mergers, while 
a population of RHs, connected with less powerful merging events, should be observable only at lower frequencies. 
A possible prototype of these RHs is that found in Abell 521 (Brunetti et al. 2008), that is indeed only barely detected at 610 MHz, and becomes clearly visible only at lower frequencies.
Being associated with less energetic merging events in the Universe, low frequency RHs are expected to be more common than ``classical'' ones, and the fraction of clusters with RHs is expected to increase 
observing at lower frequencies. In particular, it is found that the number of RHs expected in the Universe at $\simeq 100$ MHz is at least one order of magnitude larger than that expected at $\simeq 1$ GHz (Cassano et al. 2006, 2008).

Low frequency RHs are also expected to be less powerful (see Fig.\ref{fig:ussrh}.a) than classical RHs,
and thus their fraction increases with incrising the sensitivity of the survey.
For instance, preliminary calculations show that the number of RHs (over the whole sky) detectable by surveys
with typical rms $\approx 0.5$ mJy/beam and beam $=35''\times 35''$ (well reachable by LOFAR all sky surveys) is $\sim 500$ in the redshift range $0.1-0.6$ with $30\%$ of these sources visible only at low frequencies (Fig.\ref{fig:ussrh}.b). Thus the advent of the LOFAR surveys may allow to find a large number of these low frequency RHs, testing this scenario.

\section{Conclusions}

Observations reveal the presence of non-thermal emission from the ICM mainly in the radio 
band resulting from the synchrotron emission of relativistic electrons. 
The origin of the radiating electrons has always been the subject of intense debate;
however, in the last few years, some firm results have been achieved, also thanks to observational campaigns carried out with the GMRT.
It is now clearly established that RHs are relatively {\it rare} and preferentially found in massive and merging clusters: only $\sim30\%$ of a complete sample of X-ray luminous (and massive) clusters show the presence of a RH at $\approx$ GHz frequencies. 
Observations support the idea that particle acceleration due to MHD turbulence plays a role
during cluster-cluster mergers and may drive the formation of RHs.

The emerging general picture, based on a strong connection between the origin of non-thermal components in the ICM and the cluster formation process, implies an heterogeneous populations of relativistic
particles (protons, secondary and primary electrons/positrons, re-accelerated particles) mixed with the thermal ICM. The resulting non-thermal emission from galaxy clusters is a complex broad band spectrum extending from radio to gamma rays. Montecarlo calculations carried out under the hypothesis that RHs are due to turbulent re-acceleration, show that RHs are expected preferentially in massive and merging clusters, in line with present observations.
The most important expectation of the turbulent re-acceleration scenario concerns the existence of a population of low frequency RHs, missed by present radio survey at $\sim$ GHz due to their very steep spectra.
These RHs should be connected with less energetic cluster merging events, and are expected to be relatively common in the Universe. 
However, since they are also less powerful than ``classical'' RHs, deep low frequency radio surveys are necessary to discover these sources. Calculations show indeed that the number of these low frequency RHs increases with increasing sensitivity of the observations, and a considerable number of these RHs could be detected at $\sim 120$ MHz by LOFAR surveys.
The recently discovered low frequency RH in Abell 521 could be the prototype of this class of sources providing a glimpse of what LOFAR might find in next years.

\acknowledgements 
The author is very grateful to the organizing committee of the Conference and 
also thanks G. Brunetti and T. Venturi for useful comments and suggestions on the manuscript.
This work is partially supported by ASI-INAF I/088/06/0 and PRIN-INAF2007.


\end{document}